\documentclass[prl, twocolumn, showpacs, nofootinbib, amsmath,amssymb, floatfix, eqsecnum]{revtex4-1}
\pdfoutput=1
\usepackage{amsmath}
\usepackage{amssymb}
\usepackage{amsthm}
\usepackage{amsfonts}
\usepackage{comment}
\usepackage[normalem]{ulem}
\usepackage{graphicx}
\usepackage[dvipsnames]{xcolor}
\usepackage{color,framed}
\usepackage{hyperref}
\usepackage{times}
\usepackage{enumerate}
\usepackage{lipsum}
\usepackage{slashed}
\usepackage{url}
\usepackage{bbm}
\usepackage{chngcntr}
\counterwithout{equation}{section}

\hypersetup{
    colorlinks=true, 
    linktoc=all,     
    linkcolor=blue,  
}

\def \beq {\begin{equation}}
\def \eeq {\end{equation}}
\def \beqa {\begin{eqnarray}}
\def \eeqa {\end{eqnarray}}
\def \bseq {\begin{subequations}}
\def \eseq {\end{subequations}}

\begin{document}

\title{Thermalization rates and quantum Ruelle-Pollicott resonances: insights from operator hydrodynamics}

\author{Carolyn Zhang}
\affiliation{Department of Physics, Harvard University, Cambridge, MA 02138, USA}

\author{Laimei Nie}
\affiliation{Department of Physics and Astronomy, Purdue University, West Lafayette, IN 47907, USA}

\author{Curt von Keyserlingk}
\affiliation{Department of Physics, King’s College London, United Kingdom}
\date{\today}

\begin{abstract}
In thermalizing many-body quantum systems without conservation laws, such as ergodic Floquet and random unitary circuits, local expectation values are predicted to decay to their equilibrium values exponentially quickly. 
In this work we derive a relationship between said exponential decay rate $\overline{g}$ and the operator spreading properties of a local unitary evolution. A hydrodynamical picture for operator spreading allows us to argue that, for random unitary circuits, $\overline{g}$ is encoded by the leading eigenvalue of a dynamical map obtained by enriching unitary dynamics with dissipation, in the limit of weak dissipation. We argue that the size of the eigenvalue does not depend on the details of this weak dissipation (given mild assumptions on properties of the ergodic dynamics), so long as it only suppresses large operators significantly. Our calculations are based on analytical results for random unitary circuits, but we argue that similar results hold for ergodic Floquet systems. These conjectures are in accordance with existing results which numerically obtain quantum many-body analogues of classical Ruelle-Pollicott resonances [T.~Prosen J. Phys. A: Math. Gen. 35 L737 (2002), T.~Mori, arXiv:2311.10304] by studying unitary evolutions subject to weak dissipation. 
\end{abstract}

\maketitle

\textbf{\emph{Introduction.}}--- In a thermalizing many-body system, local expectation values decay towards their equilibrium ensemble averages, irrespective of most of the details of the initial state. In a system without any global symmetries, such as an ergodic Floquet and random unitary circuit (RUC), this decay is expected to be exponential \cite{prosen2002,prosen2004,keyserlingk2018,nahum2018,mori2018,lucas2019,schiulaz2019,mori2024}. This is, via linear response theory, tied to the expected exponential decay of autocorrelation functions of local operators
\begin{equation}\label{autocorr}
    |\langle O (t)O(0)\rangle-\langle O (t)\rangle\langle O(0)\rangle| \sim e^{-\bar{g}t}.
\end{equation}
Here  $\langle\cdot\rangle=q^{-L}\mathrm{Tr}(\cdot)$ is the infinite temperature ensemble expectation value, suitable for ergodic quantum systems without symmetry. $L$ denotes system size, and $q$ is the on-site Hilbert space dimension. Eq.~(\ref{autocorr}) is expected to hold until time scales proportional to $L$.

The decay rate $\bar{g}$ captures how quickly an ergodic system thermalizes, and thus locally loses information about its initial state.\footnote{Note that it is distinct from the butterfly velocity $v_B$: while $v_B$ describes how quickly operators grow apart, $\bar{g}$ describes how how slowly an operator returns, and these two rates can be very different. We discuss this point more in the Discussion and the supplemental material \cite{supp}.} Our work here concerns three connected questions: 1) How can $\bar{g}$, which describes loss of information, be understood from the underlying unitary dynamics? 2) $e^{-\bar{g}}$ is not an eigenvalue of a unitary operator such as the Floquet unitary. Can this quantity be encoded as an eigenvalue of a related operator? 3) Is $\bar{g}$ sensitive to the choice of operator $O$ or the initial state of the system? 

To answer the first question, we leverage the hydrodynamical description for operator growth introduced in Refs.~\onlinecite{nahum2018,keyserlingk2018}. In short, the information loss is closely related to the fact that, under time evolution, local operators rapidly grow in spatial support, hiding the detailed local information in the initial state into ever more delicate many-body correlators. The hydrodynamical description for operator growth can be packaged into a stochastic matrix with a leading eigenvalue 1 due to unitarity. Roughly speaking, its second leading eigenvalue can be identified with $e^{-\bar{g}}$ (there are some subtleties, which we will address later). From the hydrodynamical description, as we will show, it is also straightforward to answer question 3 in the negative, in systems without symmetry.

Question 2 was explored in classical dynamical systems, where $\bar{g}$ is related to \emph{Ruelle-Pollicott (RP) resonances}~\cite{Ruelle1986PRL, Pollicott1985AxiomA}. In classical dynamical systems, a unitary Frobenius-Perron operator $U$ generates the discrete (Floquet) dynamics on the phase space density. It was shown that after coarse-graining the phase space, $U$ becomes non-unitary, and has eigenvalues $\{e^{-\mathrm{i}\epsilon_j}\}$ within the unit circle. In certain cases, the largest eigenvalue becomes ``frozen" (independent of the coarse graining), and coincides with the leading RP resonance which sets the decay rate of autocorrelations.

\begin{figure}[tbp]
\centering
\includegraphics[width=.9\columnwidth]{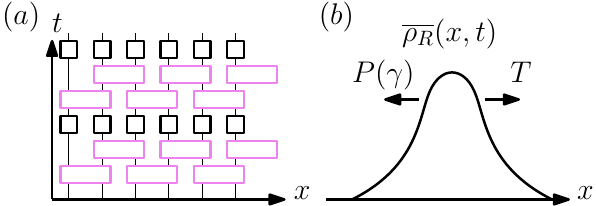} 
\caption{$(a)$ Ref.~\onlinecite{mori2024} suggested obtaining $\bar{g}$ for unitary evolution (pink rectangles) by adding a small amount of dissipation (black squares), and taking the system size $L$ to infinity before taking the dissipation rate to zero. $(b)$ In a hydrodynamical description of operator evolution, where $\overline{\rho_R}(x,t)$ describes the density of operators with right endpoint at site $x$ at time $t$, the unitary evolution is described by a stochastic matrix $T$ that tends to grow operators while the dissipation is described by a diagonal matrix $P(\gamma)$ that suppresses operators with large weight/size. }
\label{fig:circuitgamma}
\end{figure}

More recently, various authors have studied quantum analogues of RP resonances, by applying quantum versions of ``coarse-graining of phase space" \cite{prosen2002,prosen2004,mori2024,znidaric2024}. Ref.~\onlinecite{prosen2002,prosen2004} considered adding a projection $\mathbb{P}_{\ell}$ to an ergodic Floquet system, so that the Heisenberg dynamics of an operator $O$ is given by $\mathbb{P}_{\ell} \mathbb{U}_F$. Here $\mathbb{U}_F [\cdot] = U_F^\dagger [\cdot] U_F$ generates the Heisenberg dynamics under the Floquet unitary $U_F$. $\mathbb{P}_{\ell}$ is a dissipative superoperator that projects out operators with support greater than $\ell$. The resulting superoperator  $\mathbb{P}_{\ell} \mathbb{U}_F$ has eigenvalues $\{e^{-\mathrm{i}\epsilon_j}\}$ where $\{\epsilon_j\}$ are no longer all real. Numerically, the $\epsilon_j$ with smallest (negative) imaginary part, which governs the slowest decaying eigenvector, was shown to describe $\bar{g}$ obtained from the purely unitary dynamics generated by $U_F$. More recently, it was argued in Ref.~\onlinecite{mori2024} that in ergodic Floquet systems, if one adds a small amount of dephasing dissipation (see Fig.~\ref{fig:circuitgamma}) and takes the system size to infinity first before taking the dissipation rate to zero, then the resulting dissipative Floquet operator has eigenvalues $e^{-\mathrm{i}\epsilon_j}$ with an $\epsilon_j$ with smallest negative real part that matches well with the decay of autocorrelation functions (this order of limits has also been studied in the context of ``anomalous relaxation"\cite{sa2022,garcia2023,yoshimura20232}). Ref.~\onlinecite{mori2024} also gave some physical intuition for why $\bar{g}$ can be encoded as an eigenvalue of the dissipative Floquet operator: the unitary dynamics spreads an operator ballistically until it grows to a steady-state size $x^*\sim\frac{1}{\gamma}$, where dissipation and unitary evolution balance out. Then the dissipation decreases the amplitude of this operator by $e^{-\gamma x^*}\sim e^{-\bar{g}}$ where $\bar{g}\sim\mathcal{O}(1)$. 

In this work, we use the aforementioned hydrodynamical picture for operator spreading to elucidate, unify, and generalize the approaches described above for obtaining $\bar{g}$. In particular, we explain in detail how the two coarse-graining prescriptions above affect operator growth, and directly relate the modified operator growth dynamics to autocorrelation functions from unitary evolution. For the weak dissipation approach, we derive the result $x^*\sim\frac{1}{\gamma}$ explicitly, and we show why the specific order of limits $\lim_{\gamma\rightarrow 0}\lim_{L\rightarrow\infty}$ is needed to extract $\overline{g}$. More generally, we show that in addition to the two coarse-graining procedures above, \emph{any} coarse-graining procedure turning the unitary dynamics non-unitary by suppressing large operators is likely to encode $\bar{g}$ as an eigenvalue of the resulting Liouvillian.

\textbf{\emph{Hydrodynamical description of operator spreading.}}---We will first review the hydrodynamical description of operator spreading in RUCs derived in Refs.~\onlinecite{keyserlingk2018,nahum2018}. We will assume that the circuit has a brickwork geometry built out of two-site gates, acting on sites with local Hilbert space dimension $q$. Any operator can be written as a sum of operator strings $\sigma^{\nu}$, which are products of local Hermitian basis operators (for $q=2$, these are just Pauli strings)\cite{nahum2022}. Therefore, we only need to consider the evolution of operator strings. 

An operator string $\sigma^{\mu}$ generically evolves into a sum of operator strings: $\sigma^{\mu}(t)=\sum_{\nu} c_{\nu}^{\mu}(t)\sigma^{\nu}$ where $\nu$ runs over operator strings and $c_{\nu}^{\mu}(t)$ are complex coefficients. One important insight from Refs.~\onlinecite{keyserlingk2018,nahum2018} is that the evolution of an operator under a layer of random unitaries, averaged over such random unitaries, roughly depends only on the left and right endpoints of the operator (the sites farthest to the left and right that the operator acts non-trivially).
\footnote{In the present work, following Refs.~\cite{nahum2018,keyserlingk2018}, we will work with quantities averaged over an ensemble of circuits (with averaging denoted by the overline) with the conjecture that the average accurately captures the behavior of a typical circuit in the ensemble.} 

Therefore, we can group together all operator strings with the same left and right endpoints. For simplicity, we will let the original operator be at the left end of the system, so we only need to consider the movement of the right endpoints of operators.\footnote{Choosing the operator to start at the center of the system and/or keeping track of correlations between the left and right endpoints will only modify the results in this paper by subleading $\frac{1}{\mathrm{poly}(t)}$ factors.} We denote the total weight on operators with right endpoint at site $R$, averaged over circuits, by
\begin{equation}\label{recurrence}
    \overline{\rho_R^{\mu}}(x,t)=\sum_{\nu}\overline{|c_{\nu}^{\mu}|^2}\delta(\mathrm{RHS}(\nu)=x)
\end{equation}
where we sum over operator strings $\sigma^{\nu}$ with right endpoint at $x$.\footnote{We will use the same convention as in Ref.~\onlinecite{keyserlingk2018}, where each unit cell (labeled by $x$) actually has two $q$ dimensional sites.} Since a single layer of local unitaries can change any Pauli-string to any other Pauli-string with the same support, the late time evolution does not depend on the particular choice of the initial operator $\sigma^{\mu}$, and we will drop this label. It was shown in Ref.~\onlinecite{keyserlingk2018} that for RUCs, $\overline{\rho_R^{\mu}}(x,t)$ takes the form of a propagating Gaussian with a peak moving at butterfly velocity $v_B$ that depends only on $q$, where the peak width grows as $\propto\sqrt{t}$. This result was derived using the observation that $\overline{\rho_R}(x,t)$ satisfies the recurrence relation
\begin{align}
\begin{split}
    \overline{\rho_R}(x,t+1)&=2p(1-p)\overline{\rho_R}(x,t)+p^2\overline{\rho_R}(x-1,t)\\
    &+(1-p)^2\overline{\rho_R}(x+1,t)
\end{split}
\end{align}
where $p=\frac{q^2}{q^2+1}$.  This means that the right endpoints of operator strings perform a biased random walk, moving to the right with probability $p^2$, to the left with probability $(1-p)^2$, and staying on the same site otherwise. In this notation, $v_B=p^2-(1-p)^2=\frac{q^2-1}{q^2+1}$, which is smaller than the Lieb-Robinson velocity $1$ but approaches $1$ as $q\to\infty$. 

We can package the above evolution of $\overline{\rho_R}(x,t)$ as follows. For every time step, we apply the matrix
\begin{equation}\label{tpmatrix}
    T(p)=\begin{pmatrix} 1-p^2 & (1-p)^2 & 0 & \cdots\\ p^2 & 2p(1-p) & (1-p)^2 & \cdots \\ 0 & p^2 & 2p(1-p) & \cdots\\ 0 & 0 & p^2 & \cdots\\ \vdots & \vdots & \vdots & \vdots\end{pmatrix}
\end{equation}
onto the vector indexed by integer $x\in[1,L]$ with elements $\overline{\rho_R}(x,t)$. $T(p)$ is a stochastic matrix: the elements in every column sum to 1. This is because the unitarity of the evolution guarantees that $\mathrm{Tr}(\sigma^{\mu}(t)\sigma^{\mu}(t)^\dagger)$ is conserved, so $\sum_s\overline{\rho_R}^{\mu}(s)=1$ for all $t$.

The fact that $T(p)$ is a stochastic matrix means that it has an eigenvalue 1 (in this case with multiplicity 1). We conjecture that for Floquet systems, the operator dynamics is described by a similar stochastic matrix; the key properties we demand are that $T$ is quasi-diagonal and that $T_{x,y}=T_{x+1,y+1}$ for $\mathcal{O}(1)<x,y<\mathcal{O}(L)$. The latter uniformity assumption corresponds to the statement that once an operator has grown large, but is still significantly smaller than the system size, how it grows subsequently does not depend on its current size. 

\textbf{\emph{Autocorrelation decay.}}---$T(p)$ can be used to compute how autocorrelation functions decay. We will assume that under ergodic dynamics, $|c_{\nu}^{\mu}(t)|^2$ is the same for any two operators with the same left and right endpoints. In the infinite temperature state,\footnote{We give a more detailed relation between infinite temperature correlation functions and product state correlation functions in the supplemental material \cite{supp}} we get
\begin{equation}\label{sumc}
    \sqrt{|\langle\sigma^{\mu}(t)\sigma^{\mu}(0)\rangle|^2}=\sqrt{|c_{\mu}^{\mu}(t)|^2}
\end{equation}

We used the simple observation that the trace over any nontrivial operator string is zero. Note that $\overline{\rho_R}(n,t)\sim q^{2n} |c_{\mu}^{\mu}(t)|^2$ where $n$ is the extent of $\sigma^{\mu}$, so the autocorrelation function decays with the same rate as $\overline{\rho_R}(n,t)$. $\overline{\rho_R}(x,t)$ can be computed from $T(p)^t$ acting on a vector $|n\rangle$ whose only nonzero element is at site $n$. Therefore, to evaluate $\overline{\rho_R}(n,t)$, we simply compute
\begin{equation}\label{rhot}
\overline{\rho_R}(n,t)=\langle n|T(p)^t|n\rangle=\sum_k\lambda_k^t\langle n|k_R\rangle\langle k_L|n\rangle
\end{equation}
where for the second equality, we decomposed $T(p)$ into left and right eigenvectors. 

It may be tempting to immediately conclude that autocorrelations are governed by the largest eigenvalue $\lambda_0=1$ of $T(p)$. However, since $\langle n|k_R\rangle \langle k_L|n\rangle$ is exponentially small in $L$ for $k=0$ compared to $k\neq 0$, for times smaller than $\mathcal{O}(L)$, autocorrelation functions are actually governed by the subleading eigenvalues. We show in the supplemental material \cite{supp} that there are $L-1$ eigenvalues $\lambda_k=2p(1-p)(1+\cos k)$ where $k=\frac{m\pi}{L}$ for $m=1,\cdots L-1$. These eigenvalues lead to a decay rate of $(4p(1-p))^t$ (up to $\frac{1}{\mathrm{poly}(t)}$ corrections) for $t<\mathcal{O}(L)$. 

Finally at times $t\sim\mathcal{O}(L)$, the leading eigenvalue $\lambda_0=1$ dominates, causing autocorrelation functions to plateau. We can roughly estimate this plateau time by setting $4p(1-p)$ equal to the exponentially small value of $\langle n|0_R\rangle\langle 0_L|n\rangle$. For $n\sim\mathcal{O}(1)$, this gives\cite{supp}
\begin{equation}
t_{\mathrm{plateau}}\sim-2L\frac{\log p(1-p)}{\log 4p(1-p)}
\end{equation}
$t_{\mathrm{plateau}}$ grows linearly with $L$, with the coefficient going to zero as $p\to 1$. 

The above derivation of autocorrelation decay shows that there is no dependence on the observable $O$ or the initial state. The intuition is that operator growth dynamics, at long times, does not depend on the specific structure of the initial local operator $O$. Likewise, the computation of (\ref{autocorr}) in a product state is similar to the above computation; we simply need to \sout{also} sum over $|c_{\mu}^{\nu}(t)|^2$ where $\sigma^\nu$ is a operator string for which the product state is an eigenstate. It turns out that this only changes the $\mathrm{poly}(t)$ factor \cite{supp}; the exponential decay rate remains the same.

More generally, as long as $T$ is stochastic, quasi-diagonal, and uniform, we expect autocorrelation functions to decay exponentially with a constant rate, up to times of order $L$, at which point the operator has spread across the system. The decay rate is given by the subleading eigenvalues of $T$. In the above example, it is the second leading eigenvalue of $T$.\footnote{More generally there may be other eigenvalues (smaller than 1 but possibly larger than $e^{-\bar{g}}$) that, like eigenvalue $1$, only contribute once $t\sim \mathcal{O}(L)$ because they are related to boundary condition of the biased walk at $x\sim L$. See \cite{supp} for an example.} 

We will now show that coarse-graining methods described earlier all have the effect of slightly modifying the matrix $T$. For the particular $T(p)$ in Eq.~(\ref{tpmatrix}), they remove the eigenvalue 1 while, in appropriate limits, maintaining only the eigenvalues around and below? $4p(1-p)$. More generally, for systems described by other choices of $T$, the coarse-graining method results in the exponential decay observed for $t<\mathcal{O}(L)$ persisting for \emph{all} times. We will show that this implies that, for Floquet systems, the rate $\overline{g}$ is encoded as an eigenvalue of the non-unitary generator of time translations.

\textbf{\emph{Liouvillian gap: projection.}}---
The superoperator $\mathbb{P}_{\ell} \mathbb{U}_F$ has eigenvalues within the unit circle, and it was conjectured in Refs.~\onlinecite{prosen2002,prosen2004} that the leading such eigenvalue encodes $\bar{g}$. The projection $\mathbb{P}_{\ell}$ used to remove operators with size greater than $\ell$ can be easily implemented within the operator-hydrodynamical picture. We simply replace $T(p)$ by $P(\ell)T(p)$ where $P(\ell)$ is a $L\times L$ matrix $P(\ell)_{x,x}=1-\Theta(x,\ell)$ where $\Theta$ is the heaviside function. Clearly, since $P(\ell)T(p)$ is no longer stochastic, it does not have an eigenvalue $\lambda_0=1$. Notice that it is essentially a $T(p)$ on a smaller system size $\ell$, except with a modification at the $(\ell,\ell) $ matrix element that generically removes probability conservation. We show by solving the recurrence relation (\ref{recurrence}) with appropriate boundary conditions (see \cite{supp}) that $P(\ell) T(p)$ retains eigenvalues that approach $4p(1-p)$ in the thermodynamic limit. This means that autocorrelation functions decay as $(4p(1-p))^t$ even as $t\to\infty$; the projection removes the plateau at $t\sim \mathcal{O}(L)$. The eigenvectors corresponding to values near $4p(1-p)$ are all peaked near $x=\ell$. If an operator starts at size $1$, then $(P(\ell)T(p))^t$ drives it towards these dominant eigenvectors of $P(\ell)T(p)$ on an $O(\ell)$ timescale, and they dominate temporal correlations past that point. 

While the hydrodynamic description usually does not apply after $t\sim\mathcal{O}(L)$ for Floquet systems (after this point, Floquet systems differ from RUCs, because they demonstrate a ramp and plateau in autocorrelation functions), the projection keeps the operators small compared to $L$ for all times. Since operators are kept small, the operator size distribution never reaches that required for the ramp and plateau. Operator hydrodynamics should apply to operators of size $\ell$ embedded in a much larger system of size $L$, so we can continue to use operator hydrodynamics at \emph{all} times. Even in Floquet systems, we expect that the dominant eigenvectors of the dissipative evolution still correspond to operators of size near $\ell$, and a local operator will settle to them on an $O(\ell)$ timescale. 

What does this mean for the spectrum of $\mathbb{P}_{\ell}\mathbb{U}_F$? Consider the dynamics of $(\mathbb{P}_{\ell}\mathbb{U}_F)^t(O)$ for a local operator $O$. The hydrodynamical picture suggests that this operator will converge to the size $x\sim \ell$ eigenvectors over an $O(\ell)$ time-scale, and subsequently for all times decay as $\lambda^t$ for all times for some $\lambda$ (related to the aforementioned subleading eigenvalues of $T$). Therefore, for all times $t \geq O(\ell)$ we should have
\begin{equation}\label{eq:aut}
    \left|\langle O \,(\mathbb{P}_{\ell}\mathbb{U}_F)^t[O]\rangle\right|\sim \lambda^t
\end{equation}
Although the superoperator $\mathbb{P}_{\ell}\mathbb{U}_F$ has non-orthonormal left and right eigenvectors, at some $t$, the decay rate will still be dominated by the eigenvalue with the largest modulus. In other words, at some value of $t$, the eigenvalue modulus overcomes any transient features determined by wavefunction overlaps\footnote{For example, after times $t\sim\mathcal{O}(L)$, the largest eigenvalue $\lambda_0=1$ of $T(p)$ begins to dominate}. It follows that $\mathbb{P}_{\ell}\mathbb{U}_F$ needs to have a dominant eigenvalue $\lambda$. Now, notice that $\lambda$ implicitly depends on $\ell$; however we do not expect it to depend strongly on $\ell$. The idea is that autocorrelators for $\mathbb{U}_F^t$ can be represented as a sum over histories of Pauli matrices. $(\mathbb{P}_\ell \mathbb{U}_F)^t$ involves the same histories, except those involving operators larger than $\ell$ are discarded. However arguments like those in \cite{curt_backflow_1,curt_backflow_2} suggest the contributions discarded are subleading, and suppressed in $\ell$. This suggests that $\langle O \,(\mathbb{P}_{\ell}\mathbb{U}_F)^t[O]\rangle$ and $\langle O \,\mathbb{U}_F^t[O]\rangle = \langle O(t) O \rangle$ decay at similar exponential rates, so that $\lim_{\ell\rightarrow \infty} \lambda  = e^{-\overline{g}}$. This establishes that the leading eigenvalue of $\mathbb{P}_{\ell}\mathbb{U}_F$ approaches $e^{-\overline{g}}$ as $\ell\rightarrow \infty$, assuming we take the thermodynamic $L\rightarrow\infty$ limit first. 

\textbf{\emph{Liouvillian gap: dissipation.}}--- In the previous section we `coarse-grained' the unitary evolution by projecting the dynamics onto the space of small operator strings; we refer to this procedure as hard truncation. In this section we show that the main conclusions from our previous section continue to hold when we more gently  coarse-grain, by discarding longer operators through adding weak dissipation to the evolution. The effect of weak dissipation on the hydrodynamical description of operator spreading can also be implemented by a simple modification of $T(p)$. In the many body problem, interspersing unitary evolution with dissipation corresponds to stroboscopic evolution with $\mathbb{P}_\gamma \mathbb{U}_F$, with dissipative channel $\mathbb{P}_\gamma$. In the case where the dissipation is depolarising noise, for example, $\mathbb{P}_\gamma(\sigma^{\mu})=e^{-\gamma |\mu|}\sigma^{\mu}$, where  $|\mu|$ is the Pauli-weight of $\mu$ and $\gamma$ is the strength of dissipation.  In order to incorporate dissipative effects into operator hydrodynamics, we simply multiply $T(p)$ by a diagonal matrix $P(\gamma)$ which captures the effects of $\mathbb{P}_\gamma$ within the hydrodynamical picture. We take $P(\gamma)$  to be a diagonal matrix with $P(\gamma)_{x,x}=e^{-c\gamma x}$. Here, $c$ is a constant that depends on the particular channel. It does not change the result for $\gamma\to 0$, so for convenience in the following we will set $c=1$. $P(\gamma)$ captures the effects of dissipation because dissipation generically reduces operator amplitude, by an amount proportional to the operator weight\cite{schuster2023}.\footnote{For large $q$, the operator weight is approximately its extent, so we can roughly model dissipation at rate $\gamma$ by reducing the amplitude of operators by an amount exponential in their extent (the discrepancy between operator weight and operator extent can also be absorbed into the constant $c$, which does not affect the result as $\gamma\to 0$).
} 

\begin{figure}[tbp]
\centering
\includegraphics[width=.8\columnwidth]{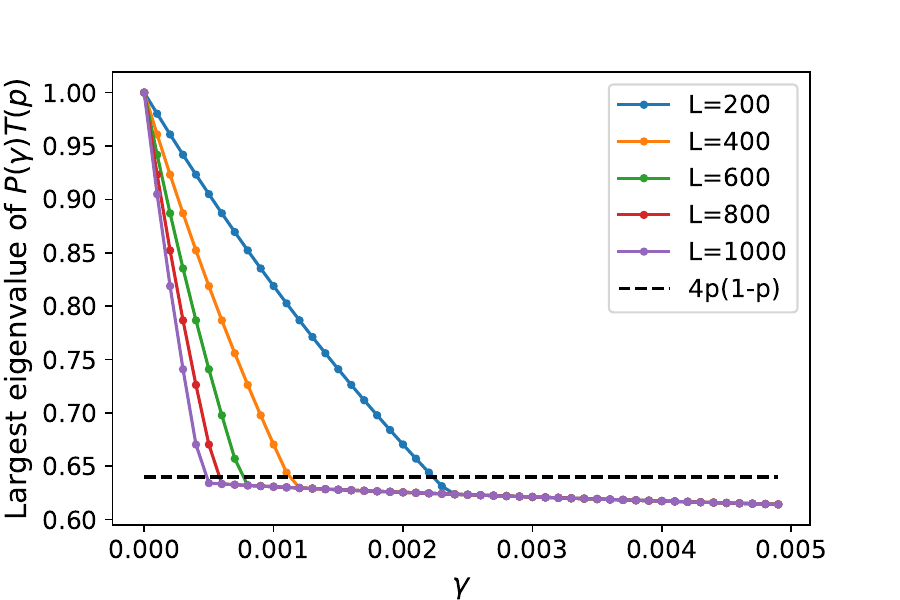} 
\caption{Leading eigenvalue of $P(\gamma)T(p)$ as a function of $\gamma$ for different system sizes, with $p = 0.8$. Black dashed line is $4p(1-p)$.}
\label{fig:mori_plot}
\end{figure}

Similar to the hard truncation, the soft truncation described by $P(\gamma)$ also removes the eigenvalue 1 of $T(p)$. Furthermore, in the above limits, $[P(\gamma)T(p)]_{x,y}\sim T(p)_{x,y}$ for $x,y\ll\frac{1}{\gamma}$. We can use our results for $P(\ell) T(p)$ to show that the eigenvalues of $P(\gamma) T(p)$ are upper bounded by $4p(1-p)$. Specifically, 
$T(p)$ has the same spectrum as the symmetric matrix $\tilde{T}(p)$\cite{supp}. Next, we can define $\tilde{T}_{\mathrm{pr}}(p)$ which has a maximum eigenvalue close to $\sim 4p(1-p)$ (according to the spectrum of $P(\ell)T(p)$) by $\tilde{T}_{\mathrm{pr}}(p)=\tilde{T}(p)-p^2|L\rangle\langle L|$. Since $P(\gamma), \tilde{T}(p)$, and $\tilde{T}_{\mathrm{pr}}(p)$ are all Hermitian and positive definite, and the spectrum of $P(\gamma)\tilde{T}(p)$ is the same as that of $P(\gamma)\tilde{T}_{\mathrm{pr}}(p)$ up to corrections $\sim\mathcal{O}(e^{-\gamma L})$, the spectrum of $P(\gamma)\tilde{T}(p)$ is bounded from above by the product of the largest eigenvalues of $P(\gamma)$ and $\tilde{T}_{\mathrm{pr}}(p)$, which is $e^{-\gamma}4p(1-p)$. We observe numerical evidence that the leading eigenvalue of $P(\gamma)T(p)$ indeed approaches $4p(1-p)$ as $L\to\infty, \gamma\to 0$ (see Fig.~\ref{fig:mori_plot}).


Furthermore, it is straightforward to see that the eigenvector 
$\overline{\rho_{R,0}}(x)$ of $P(\gamma)T(p)$ corresponding to leading eigenvalue $\lambda_0$ is peaked at size $-\frac{\log(\lambda_0)}{\gamma}$. 
The eigenstate satisfies
\begin{align}
\begin{split}\label{recurrencegamma}
    \lambda_0\overline{\rho_{R,k}}(x)&=e^{-\gamma x}\left(2p(1-p)\overline{\rho_{R,0}}(x)+p^2\overline{\rho_{R,0}}(x-1)\right)\\
    &+e^{-\gamma x}(1-p)^2\overline{\rho_{R,0}}(x+1)
\end{split}
\end{align}
Suppose that $\overline{\rho_{R,0}}(x)$ is a smooth function of $x$ (we will show that this is true in an approximation of $P(\gamma)T(p)$ later in this section; there, $\overline{\rho_{R,0}}$ has a Gaussian-like shape). Then when $\overline{\rho_{R,0}}(x)$ peaks, it flattens out, so $\overline{\rho_{R,0}}(x^*)\sim \overline{\rho_{R,0}}(x^*-1)\sim\overline{\rho_{R,0}}(x^*+1)$. If we in addition take $\gamma\to 0$, (\ref{recurrencegamma}) gives $\lambda_0=e^{-\gamma x^*}$, so $x^*=-\frac{\log{\lambda_0}}{\gamma}$. This result is consistent with the argument in Ref.~\onlinecite{mori2024}. Assuming that the leading eigenvalue of $P(\gamma)T(p)$ approaches $4p(1-p)$ in the limit $L\to\infty,\gamma\to 0$, we find that indeed operators get spread to a size $\sim-\frac{\log(4p(1-p))}{\gamma}$. Because the operator size distribution get frozen, and operators do not grow to size $\mathcal{O}(L)$, our argument from the previous section applies to this case as well. Even for Floquet systems, where $T$ may be different from $T(p)$ in (\ref{tpmatrix}), we expect the same overall picture to hold: dissipation leads to operators growing ballistically to size $x_*=O(1/\gamma)$ and becoming stuck there subsequently, with their norm decaying $e^{-\gamma x_* t}$ ever after. It follows from a similar argument to that near Eq.~\eqref{eq:aut} that the dissipative Floquet operator $\mathbb{P}_\gamma \mathbb{U}_F$ has a leading eigenvalue with modulus $e^{-\bar{g}}$.

To get intuition for the eigenvalues and eigenvectors of $P(\gamma)T(p)$, it is instructive to consider a simplified $T(p)$ obtained by dropping the upper diagonal terms. This reduces the second order recurrence relation to a first order recurrence relation, which makes it easily solvable. The resulting leading eigenvector and eigenvalues qualitatively behave in the same way as those of the full second order problem.

Expanding $T(p)$ for $p=1-\varepsilon$ in small $\varepsilon$ turns $T(p)$ into a lower triangular matrix $T(p)^{(1)}$.  $P(\gamma)T(p)^{(1)}$ is still lower triangular, and its eigenvalues are given simply by its diagonal matrix elements. In addition to an eigenstate localized at $x=L$ with eigenvalue $\sim e^{-\gamma L}$, there is a second leading eigenvalue $\lambda_1=e^{-\gamma}2\varepsilon$. Other eigenvalues are smaller, at $e^{-\gamma x} 2\varepsilon$. Therefore, if $\gamma L\ll 1$, the dominant eigenvalue corresponds to an eigenvector localized at $x=L$. However, if we take $L\to\infty$ first, and then $\gamma\to 0$, the leading eigenvalue of $P(\gamma)T(p)^{(1)}$ converges to $2\varepsilon$. For this simplified first order model, it is easy to see that autocorrelation functions also decay precisely as $(2\varepsilon)^t$. This is because if an operator grows, it cannot reduce back in size, and $2\varepsilon$ is the probability of remaining the same size. 

To get the structure of the eigenstate, we can obtain all of the ratios $\frac{\overline{\rho_R}(x)}{\overline{\rho_R}(x-1)}$ (which is simple in this case because the recurrence relation is first order) and use a telescoping product to get
\begin{equation}
\overline{\rho_R}=e^{\gamma x}\left(\frac{1-2\epsilon}{2\epsilon}\right)^xe^{-\frac{\gamma x(x+1)}{2}}\prod_{y<x}\frac{1}{1-e^{-\gamma(y-1)}}
\end{equation}
This state is roughly a Gaussian peaked at $x^*=-\frac{\log(2\epsilon)}{\gamma}$. It follows that $\gamma x^*=-\log(2\epsilon)=-\bar{g}$ for this simple model.

\textbf{\emph{Discussion.}}---There are some outstanding questions and comments. First of all, we have made the strong assumption that operator growth in ergodic Floquet many-body systems is governed by the same biased diffusive hydrodynamics as in the RUC case. This is certainly expected \cite{McCulloch_2022,tiancientdynamics} but has not been proved rigorously. One might expect that if this conjecture does hold, then the leading eigenvalue of $P_\ell \mathbb{U}_F$, and hence $\overline{g}$, should simply be determined by the butterfly velocity ($v_B$), and front diffusion constant ($D$) of the corresponding hydrodynamical process. This is actually false; we give a family of models of biased diffusion in which have identical $v_B,D$, but which predict different values of $\overline{g}$\cite{supp}. Thus $\overline{g}$ is, if it can entirely be determined by the operator hydrodynamical equations, necessarily sensitive to details of the hydrodynamics which would usually be considered irrelevant.
 
On the practical side, it would be useful to find an efficient numerical method for determining the effective stochastic matrix $T$ given an ergodic unitary evolution, perhaps using some of the ideas in Refs.~\onlinecite{McCulloch_2022,tiancientdynamics}. It would also be interesting to understand how the results are modified in the presence of symmetries, where one can study the gap of the projected dissipative Floquet operator (projecting out the conserved quantities). Finally, it would be interesting to see if an analogue of operator hydrodynamics can be applied to \emph{classical} chaos. 

\textbf{\emph{Acknowledgements}}--- We thank Sarang Gopalakrishnan, Tibor Rakovszky, Eric Schultz, Dries Sels, Sagar Vijay, and Zixia Wei for helpful conversations. C.Z. is supported by the Harvard Society of Fellows and the Simons Collaboration on Ultra Quantum Matter. CvK is supported by a UKRI FLF through MR/T040947/2 and MR/Z000297/1. This work was initiated and performed in part at the Aspen Center for Physics, which is supported by National Science Foundation grant PHY-2210452 and a grant from the Alfred P. Sloan Foundation (G-2024-22395).

\textbf{\emph{Note added}}--- 
We recently became aware of a parallel and independent work by Jacoby, Huse, and Gopalakrishnan \cite{jacoby2024} which will appear on the same arXiv posting as this manuscript. Our conclusions appear to agree where they overlap.

\bibliography{bibliography}

\end{document}


\title{Supplemental Material \\ ``Thermalization rates and quantum Ruelle-Pollicott Resonances: insights from operator hydrodynamics"}

\author{Carolyn Zhang}
\affiliation{Department of Physics, Harvard University, Cambridge, MA 02138, USA}

\author{Laimei Nie}
\affiliation{Department of Physics and Astronomy, Purdue University, West Lafayette, IN 47907, USA}

\author{Curt von Keyserlingk}
\affiliation{Department of Physics, King’s College London, United Kingdom}

\maketitle
\section{Calculation of autocorrelation decay}
After the similarity transformation 
\begin{equation}
R=\mathrm{diag}(\delta_1,\delta_2,\cdots,\delta_L)\qquad \delta_1=1\qquad \delta_n=\left(\frac{p}{1-p}\right)^{n-1}
\label{eq:similarityR}
\end{equation}
we can solve for the eigenvalues of the symmetric matrix
\begin{equation}\label{eq:tildeT}
    \tilde{T}(p)=R^{-1}T(p)R=\begin{pmatrix} (1-p)^2+2p(1-p) & p(1-p) & 0 & \cdots\\ p(1-p) & 2p(1-p) & p(1-p) & \cdots \\ 0 & p(1-p) & 2p(1-p) & \cdots\\ 0 & 0 & p(1-p) & \cdots\end{pmatrix}
\end{equation}

We use the plane wave ansatz for the (unnormalized) wavefunction 
\begin{equation}\label{psiansatz}
    \rho_{k,x}=c_1e^{ikx}+c_2e^{-ikx}
\end{equation}
Note that in the main text, we denoted eigenstates by $\overline{\rho_R}(x)$, but here we will denote them by $\rho_{k,x}$ for ease of notation. $k$ labels the particular eigenstate.

Next, we use the recurrence relation
\begin{equation}
    p(1-p)(\rho_{k,x-1}+\rho_{k,x+1})-2p(1-p)\rho_{k,x}=\lambda\rho_{k,x}
\end{equation}
Plugging in the plane-wave ansatz, we get
\begin{equation}
    \frac{\lambda-2p(1-p)}{p(1-p)}=\frac{2(c_1e^{ikx}\cos k+c_2e^{-ikx}\cos k)}{c_1e^{ikx}+c_2e^{-ikx}}=2\cos k
\end{equation}
It follows that
\begin{equation}
    \lambda=2p(1-p)+2p(1-p)\cos k
\end{equation}

Now we will derive the eigenvectors. The first recurrence relation gives
\begin{equation}
    (1-p^2)(c_1e^{ik}+c_2e^{-ik})+p(1-p)(c_1e^{2ik}+c_2e^{-2ik})=\lambda (c_1e^{ik}+c_2e^{-ik})
\end{equation}
Plugging our solution for $\lambda$ and solving for $c_1$ gives
\begin{equation}\label{c1rel}
c_1=c_2\left(\frac{p-(1-p)e^{-ik}}{(1-p)e^{ik}-p}\right)
    \end{equation}
Plugging this into (\ref{psiansatz}) gives
\begin{align}
    \begin{split}
        \rho_{k,x}&=c_2\left[e^{ikx}\left(\frac{p-(1-p)e^{-ik}}{(1-p)e^{ik}-p}\right)+e^{-ikx}\right]\\
        &=\frac{2ip\sin(kx)-2i(1-p)\sin(k(x-1))}{(1-p)e^{ik}-p}
    \end{split}
\end{align}

Picking $c_2=\frac{(1-p)e^{ik}-p}{2i}$, we get
\begin{equation}
    \psi_{k,x}\propto p\sin(kx)-(1-p)\sin(k(x-1))
\end{equation}
The last recurrence relation implements quantization of $k$:
\begin{equation}
    p(1-p)\psi_{k,L-1}+(p^2+2p(1-p))\psi_{k,L}=\lambda\psi_{k,L}
\end{equation}
Solving for $c_1$ as we did for the first recurrence relation, we find that in order for this to be consistent with (\ref{c1rel}), we must have $e^{2ikL}=1$, so $k=\frac{m\pi}{L}$ where $m=1,\cdots L-1$.

Now we would like to evaluate
\begin{equation}
    \overline{\rho_R}(n,t)=\langle n|\psi_0\rangle\langle\psi_0|n\rangle+\sum_k\lambda_k^t\langle n|\psi_{k,x}'\rangle\langle\psi_{k,x}'|n\rangle
\end{equation}
where the states $\{|\psi_{k,x}'\rangle\}$ are normalized. To normalize each state, we use
\begin{equation}
    \int_0^Ldx\langle\psi_{k,x}|\psi_{k,x}\rangle=\frac{L}{2}(1-p)^2\left(1-2p(1-p)(1+\cos(k))\right)
\end{equation}

Furthermore here we also include the normalized steady state $|\psi_0\rangle$:
\begin{equation}
    \langle n|\psi_0\rangle\langle\psi_0|n\rangle=\frac{(p/(1-p))^{2n}}{\sum_{x=1}^{L}(p/(1-p))^{2x}}=\frac{(1-p^2/(1-p)^2)(p/(1-p))^{2(n-1)}}{1-(p/(1-p))^{2L}}
\end{equation}
In the case we use $n=1$, the above simplifies because $\sin(k(n-1))=0$. In this case, we have
\begin{equation}
     \langle 1|T(p)^t|1\rangle=\frac{2}{L}\sum_k(2p(1-p)(1+\cos k))^t\frac{p^2\sin^2(k)}{\left(1-2p(1-p)(1+\cos(k))\right)} + \frac{(1-p^2/(1-p)^2)(p/(1-p))^{2(x-1)}}{1-(p/(1-p))^{2L}}
\end{equation}

Turning this sum into an integral seems to give an exponential with a subleading power law. We evaluate the integral by substituting $u=1+\cos(k):$
\begin{align}
\begin{split}
    &-\frac{2}{\pi}(2p(1-p))^tp^2\int_2^0du \frac{u^t\sqrt{2u-u^2}}{1-2p(1-p)u}\\
    &=\frac{2}{\pi}(2p(1-p))^tp^2 2^{1 + t} \sqrt{\pi}
  \Gamma[3/2 + t] F[1, 3/2 + t, 3 + t, 4p(1-p)]
\end{split}
\end{align}
where $F[\cdots]$ is the regularized hypergeometric function. We will expand the above at large $t$. Recall that
\begin{equation}
    \Gamma[3/2+t]=\frac{(2t+2)!}{4^{t+1}(t+1)!}\sqrt{\pi}
\end{equation}

We also have
\begin{equation}
    F[1,3/2+t,3+t,4p(1-p)]\sim\frac{1}{\Gamma[3+t]}+\frac{(3/2+t)4p(1-p)}{\Gamma(4+t)}+\mathcal{O}(4p(1-p))^2
\end{equation}
We will only use the first term, assuming that $4p(1-p)$ is small. Then we have
\begin{equation}
    \Gamma[3/2+t]F[1,3/2+t,3+t,4p(1-p)]\sim\frac{(2t+2)!\sqrt{\pi}}{(t+1)!(t+2)!4^{t+1}}
\end{equation}
Now we use Stirling's approximation
\begin{equation}
    n!\sim\sqrt{2\pi n}\left(\frac{n}{e}\right)^n
\end{equation}
to get
\begin{align}
\begin{split}
   \Gamma[3/2+t]F[1,3/2+t,3+t,4p(1-p)]&\sim\frac{2^{2t+2}(t+1)^{2t+2}\sqrt{2\pi(2t+2)}e^{t+1}e^{t+2}\sqrt{\pi}}{e^{2t+2}(t+1)^{t+1}\sqrt{2\pi(t+1)}(t+2)^{t+2}\sqrt{2\pi(t+2)}4^{t+1}} \\
   &\sim \frac{e}{(t+2)^{3/2}}
\end{split}
\end{align}

\begin{figure}[tbp]
\centering
\includegraphics[width=.55\columnwidth]{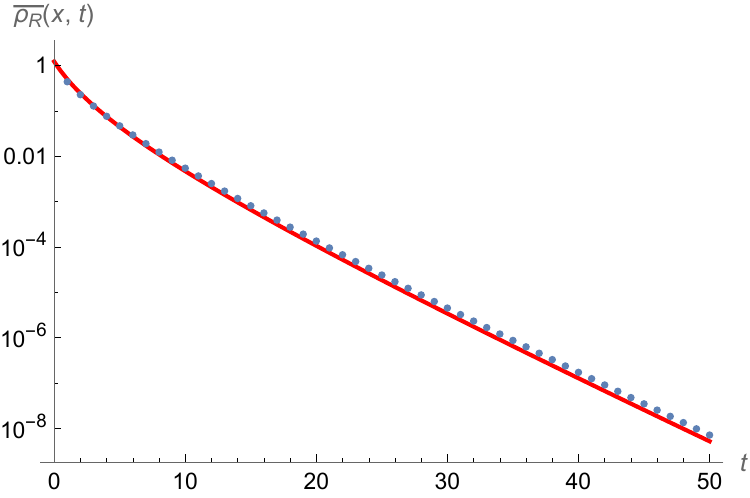} 
\caption{ The plot of $\langle x|T(p)^t|x\rangle$ computed explicitly by multiplying matrices (blue dots) vs (\ref{rhoR1}) (red line). The $x$ axis is $t$. In this calculation, $p=0.75$ and $L=28$.}
\label{fig:rhoRsmall}
\end{figure}

Therefore, we have as our final answer the following approximation:
\begin{equation}\label{rhoR1}
    \overline{\rho_R}(1,t)\sim \frac{4p^2e}{\sqrt{\pi}}\frac{(4p(1-p))^t}{(2+t)^{3/2}}+\frac{1-p^2/(1-p)^2}{1-(p/(1-p))^{2L}}
\end{equation}
In Fig~\ref{fig:rhoRsmall}, we plot (\ref{rhoR1}) and compare it to the result from numerically computing $\langle 1|T(p)^t|1\rangle$.

\begin{figure}[tbp]
\centering
\includegraphics[width=.55\columnwidth]{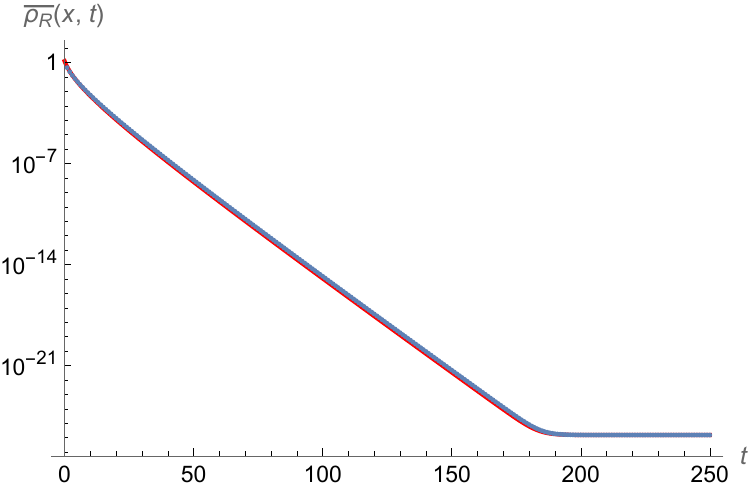} 
\caption{ The plot of $\langle 1|T(p)^t|1\rangle$ vs (\ref{rhoR1}) at much later times, including the plateau. Again, $p=0.75$ and $L=28$. The red line for cannot be distinguished from the blue dots because they closely match.}
\label{fig:rhoR}
\end{figure}

Performing a similar calculation for $\langle x|T(p)^t|x\rangle$ for $x$ finite simply shifts the coefficient by $\sim x$; the $t$ dependence remains the same.

We can roughly estimate the time at which $\langle 1|T(p)^t|1\rangle$ plateaus. Ignoring the power-law part, we get
\begin{equation}
    4p(1-p)^t\sim \frac{1-p^2/(1-p)^2}{1-(p/(1-p))^{2L}}\to t\sim-2L\frac{\log p(1-p)}{\log 4p(1-p)}
\end{equation}
This is plotted in Fig~\ref{fig:rhoR}. In the limit $p\to 1$, the coefficient of $L$ goes to zero. 

\section{Connected correlation function in a product state}
In this appendix, we will instead consider $\langle \psi|O(t)O(0)|\psi\rangle -\langle\psi| O(t)|\psi\rangle\langle\psi| O(0)|\psi\rangle$ where the expectation value is taken in a product state and $O(0)$ is a short operator string. In the fine-tuned case that  $O(0)|\psi\rangle\propto|\psi\rangle$, the $\bar{g}$ is ill defined because $\langle\psi|O(t)O(0)|\psi\rangle=\langle\psi|O(t)|\psi\rangle\langle\psi|O(0)|\psi\rangle$. 

All product states are connected to each other by a single layer of unitary gates, so it does not matter which product state we study. Suppose that the original state is a product state with every spin in the $+1$ eigenstate of $X_i$, and $O(0)=Z_i$. Then $\langle\psi| O(0)|\psi\rangle=0$ and $\langle \sigma^{\nu} Z_i\rangle=0$ unless $\sigma^{\nu}=Y_i\prod_{j} X_j$ or $Z_i\prod_{j}X_j$ for some product over spins $j$. Let us denote this subset of operator strings by $\{\sigma_{n'}\}$. Then
\begin{equation}
   | \langle \psi|O(t)O(0)|\psi\rangle-\langle\psi| O(t)|\psi\rangle\langle\psi| O(0)|\psi\rangle|^2 = \sum_{n',m'}c_{n'}(t)c_{m'}^*(t)\sim\sum_{n'}|c_{n'}(t)|^2
\end{equation}
where we drop terms with $n'\neq m'$ assuming that they come with random signs that cancel. Note that for the set of operator strings acting on an interval of length $l$ (not necessarily acting nontrivially on every site within the interval), the probability that the string is of the form mentioned above is $\sim \frac{1}{q^{2l}}$ (recall that in the convention we use, a single unit cell labeled by $x$ actually has two $q$ dimensional spins). This is because there are $q^{2l}$ strings in the interval of the form $Y_i\prod_jX_j$ or $Z_i\prod_jX_j$ but $q^{4l}$ operators in total. If we assume that for a given length, all operator strings occur with the same probability, i.e. $\overline{|c_{n'}(t)|^2}=h(l,r,t)$ where $r,l$ are the right and left and points, we get $\langle\psi| O(0)|\psi\rangle=0$, then we just need to integrate over $\overline{\rho_R}(x,t)$ with a weight $q^{-2x}$.

This is easiest to do if we consider a slightly different setup where we allow the right endpoint to travel to negative values (or we can place the right endpoint of the operator at the middle of the system rather than at site 1). Denote $\overline{\rho_R}(x,t)$ computed in this way by $\overline{\rho_R}'(x,t)$. This can be obtained from the standard generating functional method as
\begin{equation}
    \overline{\rho_R}'(x,t)=\frac{q^{2(t+x)}}{(1+q^2)^{2t}}{2t \choose t+x}
\end{equation}
For finite $x$, expanding in large $t$ gives the expected $4p(1-p)^t$ with a $\frac{1}{\sqrt{t}}$ subleading power law decay rather than the $\frac{1}{t^{3/2}}$ derived above. The amplitude is higher because there are more trajectories for returning to the original site (set to be zero) if the right endpoint can go to negative values. 

To compute $\sum_{n'}|c_{n'}(t)|^2$ we sum over $\overline{\rho_R}(x,t)$ for all $x>0$ with a factor of $q^{-2x}$. This gives $\sum_x\overline{\rho_R}'(x,t)q^{-2x}\sim (4p(1-p))^t$. The summation that differentiates the product state from the infinite temperature state only changes the $\mathrm{poly}(t)$ factor. We expect a similar calculation performed using the approach in the previous appendix (which only counts trajectories of the right endpoint that stay above $x=0$) would also only change the $\frac{1}{t^{3/2}}$ factor.

\section{Analytical estimation of the leading eigenvalue of $P(\ell)T(p)$}

In this appendix, we show that for generic hard truncations such as $P(\ell) T(p)$ described in the main text, the spectrum of $P(\ell)T(p)$ consists of eigenvalues bounded above by $4p(1-p)$ and that the leading eigenvalue is within $\mathcal{O}\left(\frac{1}{L^2}\right)$ of $4p(1-p)$.

We consider the $\ell\times\ell$ matrix
\begin{align*}
\begin{pmatrix}
a+b & a & 0 & 0 & \cdots\\
c & b & a & 0 & \cdots\\
 0 & c & b & a & \cdots\\
 0 & 0 & c & \cdots & a\\
 0 &  0 &  \cdots& c & b+c'
\end{pmatrix}
\end{align*}
where we use the short hand $a=(1-p)^2,b=2p(1-p),c=p^2$ with $1>p>\frac{1}{2}$ for RUCs. $c'=c$ for $T(p)$ on a system on size $\ell$ and $c'=0$ for the upper left block of $P(\ell) T(p)$. $c'=c$ gives a stochastic matrix while $c'=0$ does not. We may have different $a,b,c$ for ergodic Floquet systems. 

The eigenvectors satisfy
\begin{align}
\begin{split}\label{recrel}
0 & =(\tilde{\lambda}-a)\rho_{1}-a\rho_{2}\\
0 & =\tilde{\lambda}\rho_{x}-a\rho_{x+1}-c\rho_{x-1},\qquad x\in\{2,\ldots,\ell-2\}\\
0 & =(\tilde{\lambda}-c')\rho_{\ell}-c\rho_{\ell-1}
\end{split}
\end{align}
where $\tilde{\lambda}=\lambda-b$. To solve this difference equation we first solve the characteristic equation $ay^2-\tilde{\lambda}y+c=0$. This has two roots
\begin{equation}
    y_{\pm}=\frac{\tilde{\lambda}\pm\sqrt{\tilde{\lambda}^2-4ac}}{2a}
\end{equation}
which are distinct if $\tilde{\lambda}^2-4ac\neq 0$. Our first task will be to show that there are no roots in the interval $1\geq\lambda>4p(1-p)$ for $c'$ not fine-tuned. We will then show that there is a root within $\mathcal{O}\left(\frac{1}{\ell^2}\right)$ of $4p(1-p)$.

Suppose that $1\geq\lambda>4p(1-p)$. Then we write
\begin{equation}
    \rho_x=\alpha_+y_+^{x-1}+\alpha_-y_-^{x-1}
\end{equation}

Note that all non-trivial solutions to the recurrence relation have $\rho_1\neq 0$. Up to overall normalization of the state, we can assume that $\rho_x=1$, so $\alpha_{\pm}=(1-\alpha),\alpha$. The first recurrence relation in (\ref{recrel}) means that
\begin{equation}
    \left(\tilde{\lambda}-a\right)-a(\alpha y_++(1-\alpha)y_-)=0\to\alpha=\frac{ay_-+a-\tilde{\lambda}}{a(y_--y_+)}
\end{equation}
Note that $\alpha$ does not scale with $\ell$ and is finite and nonzero for $1\geq \lambda>4p(1-p)$ 
If $\lambda>4p(1-p)$ by an $\mathcal{O}(1)$ amount, then $y_+>y_->0$, so $y_+^{x-1}$ dominates at large $x$. Therefore,
\begin{align}
\begin{split}
\rho_{\ell} & =\alpha y_{+}^{\ell-1}\left(1+e^{-O(\ell)}\right)\\
\rho_{\ell-1} & =\alpha y_{+}^{\ell-2}\left(1+e^{-O(\ell)}\right)
\end{split}
\end{align}

Plugging this into the last relation in (\ref{recrel}), we find that
\begin{equation}
    \left(\tilde{\lambda}-c'\right)y_+-c=e^{-\mathcal{O}(\ell)}
\end{equation}
which does not have any solutions for $\lambda>4p(1-p)$ except along a fine-tuned line $\lambda(c')=\frac{c}{y_+}+c'+b$ (neglecting terms $\mathcal{O}\left(e^{-\ell}\right)$). For example, for the hard projection $c'=0$, we get
\begin{equation}
    \tilde{\lambda}y_+-c=\frac{\tilde{\lambda}^2-4ac+\tilde{\lambda}\sqrt{\tilde{\lambda}^2-4ac}+2ac}{2a}\geq c>0
\end{equation}
We used the observation that the first two terms in the numerator are positive when $\lambda>4p(1-p)$. It follows that for sufficiently large $\ell$, there is no solution for $c'=0$, and there are no eigenvectors with eigenvalue $\lambda>4p(1-p)$.

If instead $c'=c$, then we need to solve
\begin{equation}
    (\tilde{\lambda}-c)\rho_{\ell}-c\rho_{\ell-1}=0\to (\tilde{\lambda}-c)y_+-c=e^{-\mathcal{O}(\ell)}
\end{equation}
This equation only has the solution $\lambda=1$ for $\lambda>4p(1-p)$. Note that when $\lambda=1$, $1-\alpha=0$, so this solution also holds for finite $\ell$.

Next, we will show that there are roots approaching $4p(1-p)$ from below, in the case $c'\neq c$ (for $c'=c$, we already know that there are roots near $4p(1-p)$ from the exact solution for the eigenvalues of $T(p)$). It suffices to show that
\begin{equation}
    f(\lambda)=\tilde{\lambda}\rho_{\ell}-c\rho_{\ell-1}
\end{equation}
changes sign in the vicinity of $\lambda=4p(1-p)$, so it passes through zero. Let $\lambda=4p(1-p)-\epsilon=2b-\epsilon$ for $\epsilon>0$. In this case,
\begin{equation}
    y_{\pm}=\frac{b-\epsilon\pm i\sqrt{(2b-\epsilon)\epsilon}}{2a}=re^{i\theta}
\end{equation}
where $r=\sqrt{\frac{c}{a}}$ and $\theta=\arctan\left(\sqrt{\frac{(2b-\epsilon)\epsilon}{(b-\epsilon)^2}}\right)$. Now $y_{\pm}$ have equal modulus so $y_+^{\ell}$ no longer dominates the expression for $\rho_{\ell}$. From the first recurrence relation in (\ref{recrel}), we now have
\begin{equation}\label{alphaphi}
    \alpha=\frac{are^{-i\theta}+a-(b-\epsilon)}{ar\left(e^{-i\theta}-e^{i\theta}\right)}=|\alpha|e^{i\phi}
\end{equation}

Let $g(\epsilon)=f(2b-\epsilon)$. Then
\begin{align}
\begin{split}
\tilde{g}(\epsilon)\equiv g(\epsilon)\frac{r^{2-\ell}}{2|\alpha|} & =\frac{(b-\epsilon)r}{2|\alpha|}(\alpha e^{(\ell-1)i\theta}+(1-\alpha)e^{-(\ell-1)i\theta})-(\alpha e^{(\ell-2)i\theta}+(1-\alpha)e^{-(\ell-2)i\theta})\\
 & =\frac{(b-\epsilon)r}{2|\alpha|}(\alpha e^{(\ell-1)i\theta}+\alpha^{*}e^{-(\ell-1)i\theta})-(\alpha e^{(\ell-2)i\theta}+\alpha^{*}e^{-(\ell-2)i\theta})\\
 & =(b-\epsilon)r\cos((\ell-1)\theta+\varphi)-\cos((\ell-2)\theta+\varphi)
\end{split}
\end{align}

Expanding around small $\epsilon$, we get $\theta=\sqrt{\frac{2\epsilon}{b}}$, so we need $\epsilon\sim\frac{1}{\ell^2}$ for $\ell\theta\sim\mathcal{O}(1)$. Plugging this expression for $\theta$ into (\ref{alphaphi}) and again expanding in small $\epsilon$, we get $\phi=\frac{\pi}{2}+\frac{1}{b-1}\sqrt{\frac{2\epsilon}{b}}=\frac{\pi}{2}+\frac{1}{b-1}\theta$. Then using $\cos(x+\pi/2)=-\sin(x)$, we get
\begin{equation}
    \tilde{g}(\epsilon)=(b-\epsilon)r\sin\left((\ell-2)\theta+\frac{\theta}{b-1}\right)-\sin\left((\ell-1)\theta+\frac{\theta}{b-1}\right)
\end{equation}
Suppose that $\epsilon=\frac{\psi}{\ell^2}$ where $\psi$ is $\mathcal{O}(1)$. Then the above simplifies to
\begin{equation}
    \tilde{g}(\epsilon)=(1-2p^2)\sin\left(\sqrt{\frac{\psi}{p(1-p)}}\right)+\mathcal{O}\left(\frac{1}{\ell}\right)
\end{equation}
Therefore, $\tilde{g}(\epsilon)$ passes through zero at $\frac{\psi}{p(1-p)}=\pi\to \psi=p(1-p)\pi^2$, which is the largest root. The second largest root occurs near $\psi=4p(1-p)\pi^2$.

\section{Dependence of $\bar{g}$ on $v_B,D$}
In this section, we will show that $\bar{g}$ is not fully determined by just $v_B$, or even $v_B$ together with the diffusion constant $D$. We present a minimal example of a family of biased random walks with the same $v_B$ and $D$ but with different $\bar{g}$.

Consider the lower triangular stochastic matrix
\begin{equation}
T(p,\epsilon)=\begin{pmatrix} 1-p-\epsilon & 0 & 0 & 0 & 0 & \cdots\\
\frac{p}{2}+3\epsilon & 1-p-\epsilon & 0 & 0 & 0 & \cdots\\ \frac{p}{2}-3\epsilon & \frac{p}{2}+3\epsilon & 1-p-\epsilon & 0 & 0 & \cdots\\ \epsilon & \frac{p}{2}-3\epsilon &\frac{p}{2}+3\epsilon & 1-p-\epsilon & 0 & \cdots \\ 0 & \epsilon & \frac{p}{2}-3\epsilon & \frac{p}{2}+3\epsilon & 1-p-\epsilon & \cdots\\ \vdots & \vdots & \vdots & \vdots & \vdots & \vdots \end{pmatrix}
\end{equation}
which is nonnegative if $0\leq\epsilon\leq \mathrm{min}(\frac{p}{6},\frac{1-p/2}{3})$. The leading eigenvalue of $P(\gamma)T(p)$ as we take $L\to\infty$ followed by $\gamma\to 0$ is $1-p-\epsilon$ (this is easy to see from the fact that $T(p,\epsilon)$ is lower triangular). However the recurrence relation describes a biased random walk with velocity and diffusion constant that is independent of $\epsilon$. Specifically, $v_B=\frac{3p}{2}$ while $D=\frac{5p}{4}$. $\epsilon$ shows up as a coefficient of the $\partial_x^3$ term. 

It is also straightforward to construct simple examples where the leading eigenvalue of $P(\gamma)T(p)$ (where $T(p)$ is the stochastic matrix in the main text) does not coincide with the second largest eigenvalue of $T(p)$. For example, we can consider the lower triangular matrix above, and make the second to last entry on the diagonal $1-p$ rather than $1-p-\epsilon$. We can modify the rest of the column so that it remains stochastic. In this case, the second largest eigenvalue of $T(p)$ is $1-p$, bt this eigenvalue is suppressed almost as strongly as the eigenvalue 1; it will only appear at $t\sim\mathcal{O}(L)$.

\section{Numerical results on the eigenstates of $P(\gamma)\tilde T(p)$ and $P(\gamma)T(p)$}
\begin{figure}[htbp]
\centering
\includegraphics[width=.45\columnwidth]{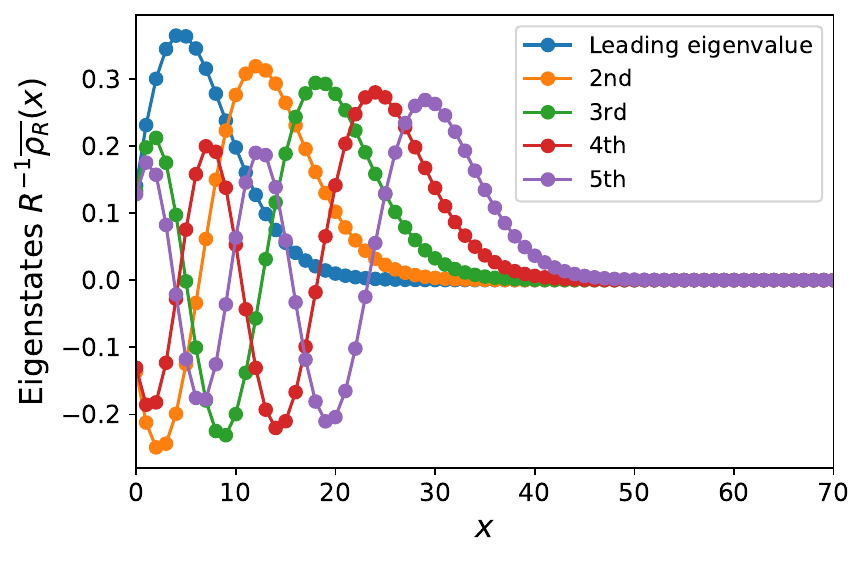}
\hspace{1cm}
\includegraphics[width=.45\columnwidth]{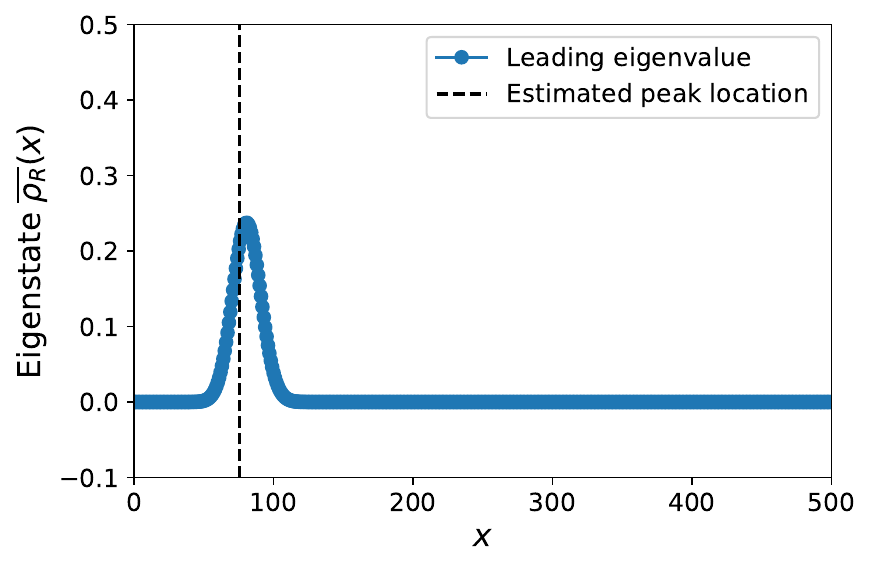}
\caption{Left: eigenstates of $P(\gamma) \tilde T(p)$, $R^{-1}\overline{\rho_R}(x)$, for the first few eigenvalues. $\overline{\rho_R}(x)$ are the eigenstates of $P(\gamma)T(p)$, and the similarity transformation $R$ is defined in~\eqref{eq:similarityR}. Right: eigenstate of $P(\gamma) T(p)$, $\overline{\rho_R}(x)$, for the leading eigenvalue. In both plots $p=0.8, L = 500, \gamma = 3/L = 0.006$ (we only plot to $x=70$ on the left). The leading eigenvalue is $\lambda_1 = 0.621$, close to our theoretical estimation $e^{-\gamma} 4p(1-p) = 0.638$. The peak location on the right plot is close to our estimation $-\log (\lambda_1) / \gamma = 75.381$ (black dashed line) as well.}
\label{fig:Eivenvecs}
\end{figure}

Here we present numerical solutions of the eigenstates of $P(\gamma)\tilde T(p)$ and $P(\gamma) T(p)$. As shown in the left plot of Fig.~\ref{fig:Eivenvecs}, most of the eigenstates of $P(\gamma)\tilde T(p)$, which is the dissipative matrix after the similarity transformation (see~\eqref{eq:tildeT}), have oscillating behaviors. Because the eigenstates describe probability density distributions, we discard those that are not non-negative and only keep the one that belongs to the largest eigenvalue. The right plot of Fig.~\ref{fig:Eivenvecs} shows the spatial profile of this non-negative eigenstate with similarity transformation added back on, i.e. the eigenstate of the original $P(\gamma) T(p)$. Its peak location is close to our analytical estimation (black dashed line).

\bibliography{bibliography}